# A Tool to Automate the Sizing of Application Process for SOA based Platform

Debajyoti Mukhopadhyay
Department of Information Technology
Maharashtra Institute of Technology
Pune, India
debajyoti.mukhopadhyay@gmail.com

Juhi Jariwala
Department of Information Technology
Maharashtra Institute of Technology
Pune, India
juhi.jari@gmail.com

Payal Innani
Department of Information Technology
Maharashtra Institute of Technology
Pune, India
innani.payal16@gmail.com

Sheetal Bablani
Department of Information Technology
Maharashtra Institute of Technology
Pune, India
sheetal.bablani.08@gmail.com

Sushama Kothawale
Department of Information Technology
Maharashtra Institute of Technology
Pune, India
sushkothawale@gmail.com

*Abstract*— **SOA (Service Oriented Architecture) is a loosely-coupled architecture designed to tackle the problem of Business/Infrastructure alignment to meet the needs of an organization. A SOA based platform enables the enterprises to develop applications in the form of independent services. To provide scalable service interactions, there is a need to maintain service's performance and have a good sizing guideline of the underlying software platform. Sizing aids in finding the optimum resources required to configure and implement a system that would satisfy the requirements of BPI(Business Process Integration) being planned. A web based Sizing Tool prototype is developed using Java APIs(Application Programming Interface) to automate the process of sizing the applications deployed on SOA platform that not only scales the performance of the system but also predicts its business growth in the future.**

## I. INTRODUCTION

In today's e-world, where companies and organizations carry out most of their business over the internet, software architectures attempt to deal with the increasing levels of complexity. As the level of complexity continues to ascend, traditional architectures do not seem to be capable of dealing with the current problems such as the need to respond quickly to new requirements and allow better and faster integration of applications. As web technologies and related business needs evolve, especially isolated service requests/components, it puts a new demand on the software architecture being used. SOA being a platform independent architecture, it provides viable working solution to implement dynamic e-business. It enables distinct software applications to exchange information and communicate with other applications in the network without human interaction and without the need to make changes to the underlying software program itself.

A service-oriented architecture[1] is essentially a collection of services, among which the communication can involve either by simple data passing or it could involve two or more services coordinating some activity, requiring means of connecting services to each other. The first service-oriented architecture in the past was with the use DCOM or Object Request Brokers (ORBs) based on the CORBA specification[2].

To understand service-oriented architecture it is important to have a clear understanding of the term service. A service[3] is a function that is well defined, self-contained, and does not depend on the context or state of other services. Figure 1 illustrates a basic service-oriented architecture wherein a service consumer sends a request to a service provider and the service provider replies back with a response. Communication via messages promotes interoperability, and thus provides adaptability benefits, as messages can be sent from one service to another without the consideration of how the service handling those messages has been implemented. SOA makes it easy for computers connected over a network to cooperate with one another. Every computer can run an arbitrary number of services, and each service is built in a way that ensures that the service can exchange information with any other service in the network without the need of making changes in the base program.

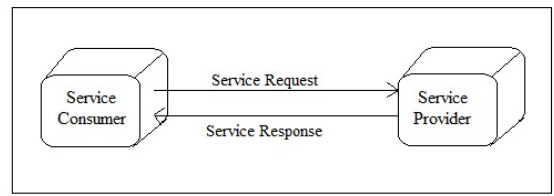

Fig 1. Basic Service Oriented Architecture

The key characteristic of SOA is that these independent services with standard interfaces can communicate with other services in a way, without the service having foreknowledge of the calling service and of the underlying platform implementation. SOA based platforms are developed to facilitate the integration of business processes deployed on various enterprise applications. One important requirement is to ensure that the deployed system satisfies the performance objective. To help the architects and business analysts control the cost of a BPI solution, it is often required that the resource consumption is estimated before the system is developed and deployed. That is, sizing of the platform needs to be carried out. Considering the limitations of the existing sizing tools available and to automate the sizing process, a web based sizing tool prototype is proposed. It determines the required hardware along with its topology before deploying the applications, taking into considerations the best practices.

Section 2 explains the literature survey carried out. Section 3 provides a complete view of the proposed tool and its development in phases. Section 4 describes the implementation details and results. Section 5 covers the practical applications of the prototype. Sections 6, 7 present the future scope and conclusions respectively.

## II. EXTENSIVE TECHNICAL RESEARCH

Following the global economic growth, firms carrying out their business over the web look for innovative ways to cut down the technical costs and maximize its value, in order to acquire a competitive hold on the IT(Information Technology) market. Growing acceptance of pioneering technologies makes terms like SOA, web services, sizing, etc. big buzzwords in IT. Owing to their importance, the technical research phase led to the study of following topics:

1. Web Services and its applications

Web services are not tied to any operating system or programming language. Along with supporting convenient and on-demand communication, they provide a strong interface for collection of operations being accessed on the network. It has applications in cloud computing, wherein a Service Consumer can choose "a-right" service from a group of similar web services.[4] Web services are also used in query optimization techniques to come up with faster and efficient algorithms to retrieve data from databases.[5]

2. Application of SOA

The telecommunication industry focuses on delivering the best quality services to its users. To keep the system up and running at its best, capacity planning of the services and of the platform is carried out to be able to handle increasing incoming transaction requests with allocated services[6].

3. Existing Sizing Tools and Guides

Presently, there are sizing tools and sizing guides available in the market. Some being precise to solving a particular domain problem whereas some being specific to certain products only.

a. HP's(Hewlett-Packard) Performance and Sizing Guide[7] provides information and recommendations about designing and configuring the UNIX environment to run its Administration UI(User Interface). Although being API based, some of its operations do not rely on the API.

b. Performance Tuning and Sizing Guide for SAS Users[8] gives basic understanding of how to analyze and apply tuning changes to SAS applications running on SUN UltraSPARC hardware platform.

c. Intel's Server Sizing Tool[9] provides its customers with a sizing solution in the form of two-socket servers or four-socket servers based on various workload conditions in their ERP(Enterprise Resource Planning) environment. However, being limited to a particular environment, it may not be able to size servers belonging to other environments.

The objective of the sizing prototype discussed in this paper is to present a generalized, API based tool for SOA platforms constructed using a lightweight UI framework.

## III. PROPOSED WORK

Currently the process of sizing a SOA platform is performed by referring the available documents and related paper work. But many a times, customers are unable to understand it or do not have access to such legal documents. Also, filtering what data is useful for the end-user from the historical data is a tedious task. Having an automated process would take the sizing process one step up on the sizing ladder and overcome the above mentioned restrictions. Therefore, we have come up with a sizing tool prototype to provide meticulous hardware recommendations for the applications and services deployed on SOA based platforms. This tool comes up with sizing suggestions and implementation details according to which a durable and robust system can be developed. Also resources can be utilized in a competent and well-organized way. This in the long run ensures lengthening life of the hardware as well as reducing maintenance and repairs cost. The sizing model is constructed in three phases, testing and analysis phase, API generation phase and UI(User Interface) phase. Each of these phases is described below:

*A. Performance Testing and Analysis phase*

Simple applications are deployed on the AMX platform and tests are conducted using different performance testing tools. These tests are performed to study how the system performs when load on the deployed applications is varied. Meaningful data is collected from these tests and analyzed.

*B. API generation phase*

The analyzed data is then used to determine a relation between the performance parameters in the form of equations. An algorithm is built using these equations. Further, this algorithm forms a basis of a generalized Java based API. APIs can be easily modified and reused.

*C. Frontend UI phase*

To provide the end user an easy and user friendly way of performing the sizing task, a lightweight UI is constructed to give a visual feel of their system's performance.

Following diagram shows the visualization of the sizing tool prototype:

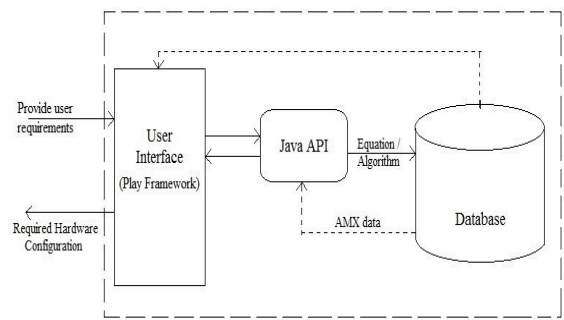

Fig 2. Visualization of Sizing Tool Prototype

## IV. IMPLEMENTATION

*A. Experimental Setup:*

TIBCO ActiveMatrix(AMX) is a platform provided by TIBCO for developing and deploying distributed SOA based applications. Using this platform, enterprises can rapidly design, implement, and test applications, deploy them to their operating environment of choice and monitor and manage the applications end-to-end. There are various parameters like type of service, payload(request and response), concurrent users, throughput, etc. that need to be taken into consideration while sizing this JAVA based platform. To obtain relation between these parameters and to monitor how system resources are utilized if these parameters are assorted, various performance related load tests are conducted on this platform. Initially applications are deployed on the platform. Each application is assigned an individual port number where it can receive requests and process its task accordingly. Various tools are used to generate load tests on those applications. During load tests performance parameters like CPU utilization, heap, etc. required by services are observed of the respective JVM.

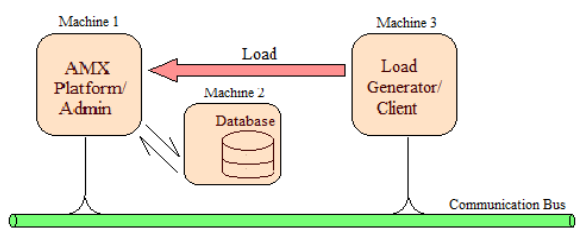

Fig 3: Experimental and Testing Setup

*B. Testing Methodology*

This sizing framework aims to put forth its sizing summary in the form of 3 standard system recommendations heavily used in production environments, namely, medium system, large system and performance lab system. The hardware configuration of these systems is as follows:

Table 1. Hardware Configurations

| Type of System | No. of Processors | No. of Cores | Frequency (GHz) | RAM (GB) |
|---|---|---|---|---|
| Medium | 2 | 4 | 3.07 | 32 |
| Large | 2 | 8 | 3.07 | 64 |
| Performance Lab | 2 | 12 | 3.07 | 64 |

All tests and experiments are conducted on the performance lab system i.e. on a logical 24 core machine. Considering several variable parameters, the task of deploying services and making them run efficiently is very complex. Tests are performed by simulating real customer scenarios that meet performance requirements. To generate understandable results, analysis is conducted in the following steps:

Step 1: Same load tests are carried out on each of the three systems mentioned above and depending on results obtained from these tests relation for CPU utilization and memory usage between these systems is derived for medium system and large system in terms of performance lab system.

Step 2: All the further load and stress tests are carried out on the performance lab system and values for medium and large system are generated according to the relation generated in step 1.

Step 3: Load tests are performed to find relation between all input parameters. Scalability aspects involving the memory and CPU capability and features are also tested.

Step 4: Using results of all load tests and relation between parameters obtained in step3, graphs are plotted and equations are derived for hardware resources required by services.

Step 5: The equations obtained are fine tuned by extrapolating the values and further validating them. This helps in generating more accurate results.

*C. Knapsack's Algorithm*

The 0/1 Knapsack Problem[10] is stated as follows:

Given a set of 'n' kinds of items labelled from 1 to n, having a weight 'w' and value 'v' associated with each item respectively, determine the maximum number of items to be included in a collection so that the total weight is less than or equal to a given limit, say 'W' and maximize the total value as much as possible.

This can mathematically be represented as bellow:

$$\text{maximize} \sum_{j=1}^{n} v_j x_j$$

$$\text{subject to} \sum_{j=1}^{n} w_j x_j \leq W$$

$$x_j \in \{0,1\}$$

The knapsack's problem is further classified into the Bounded Knapsack Problem(BKP) and the Unbounded Knapsack Problem(UKP). In BKP, 'x' can take up values ranging from 0 to some constant whereas in UKP, there is no restriction on the value of 'x' except that it should be a non-negative integer.

In order to harness the hardware and software resources in the most efficient way, we chose the Unbounded Knapsack algorithm and implemented it in the following manner:

```
Let current_machine_cpu=0,current_machine=1
for i from 1 to number_of_services
do
  if current_machine_cpu+cpu_of_i_th_service<W
    deploy service on current_machine
    current_machine_cpu+= cpu_of_i_th_service
  else
    for j from i+1 to number_of_services
    do
      if current_machine_cpu+cpu_of_j_th_service<W
        deploy service on current_machine
        current_machine_cpu+= cpu_of_j_th_service
      end if
    end for
  end if
  if current_machine_cpu>=W
    distribute services among the nodes on that machine
    switch to next machine
  end if
end for
```

The above algorithm can mathematically be represented as shown below:

$$\text{maximize} \sum_{j=1}^{n} v_j x_j$$

$$\text{subject to} \sum_{j=1}^{n} w_j x_j \leq W$$

where,
$x_j$ is $j^{th}$ service
$v_j$ is type of $j^{th}$ service
$w_j$ is CPU utilized by $j^{th}$ service
W is maximum allowed CPU% utilization per machine

### D. Sizing API

To give the best and optimum hardware recommendation to the end user, this tool summarizes its solution using sizing API (Application Programming Interface). This API is developed using the Knapsack algorithm and mathematical equations that forms the backbone of this web based tool. As there are new hardware configurations continuously evolving in the industry, it will be required to make changes in standard server configurations considered (as explained in section 3.B.) for the output of our tool. With sizing API, these changes can be envisaged easily without the need to change the code and hence it can be extended to be the underlying computation for sizing of other SOA based products. The following code snippet represents one part of the API constructed:

```
for(j=1;j<=StandardConfiguration.large_system
    if(remaining_services>0){
        services=large_system_services_per_node
        remaining_services--;
    }
    else
        services=large_system_services_per_node

    large_system_total_nodes--;
    distributed_architecture_max_nodes--;
    Node n=new Node(services,host_count);
    m.node.add(n);

    if(distributed_architecture_max_nodes==0)
        Host h=new Host(i);
        m.host.add(h);
        host_count++;
        distributed_architecture_max_nodes=5;
```

Fig 4.Sizing API

### E. System Workflow

Customer priorities vary depending on the criticality of applications. While some applications are critical because of their payload, some are critical due to large number of concurrent users. However, we need to consider the combined effect of these factors while designing the sizing solution. Therefore the input parameters are broadly categorized into two sections, namely, deployment time inputs and runtime inputs. Deployment time inputs include number of services to be deployed, implementation and binding type of those service(s). Whereas, the Runtime input include workload type, concurrency, throughput and payload of the service(s). The user is also given the option to enter his choice of architecture i.e. either Single or Distributed. These inputs are classified into two fragments mainly to furnish the users with two levels of sizing outputs. The output of the first level gives a deployment time sizing suggestion to the customers. That is, the amount of hardware required to only deploy the specified number of services. In the next level, the output suggested by the tool also considers the runtime values along with the deployment values. The tool provides a summary page which collates all the inputs given by the customer, and enables them to trace back and change the values or options if required. Following flow chart shows working of the system:

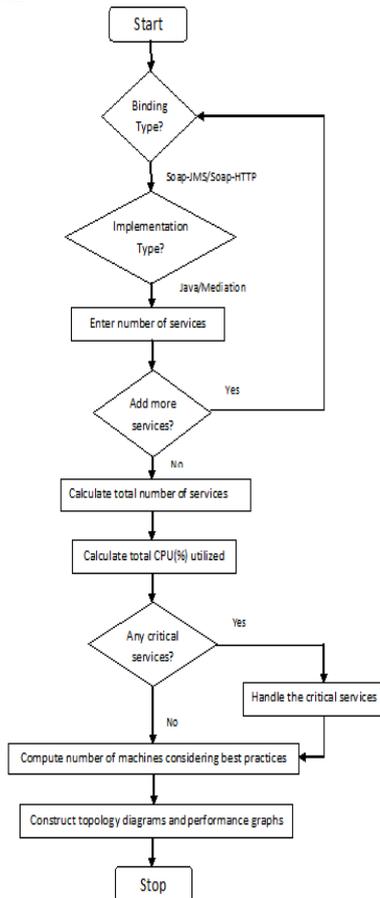

Fig 5. Flowchart of system workflow

*F. Results*

The sizing tool adheres to a mathematical modelling approach which delivers a solution based on the equations obtained from testing and analysis phase. The results generated are in the form of topology diagrams, performance graphs and infrastructure diagrams for all the 3 system recommendations i.e. medium system, large system and performance lab system. This kind of comparison gives a pictorial view to the customers suggesting which system implementation they should adopt that would best maximize their available resource utilization. The user also has an added benefit of downloading the summary report. This comprehensive report gives all the required information in detail to set up the system. For architects and analysts, such reports come handy in determining how resources are to be allocated and how much of the hardware will be required to build the system.

Along with this, the tool also suggests the kind of architecture that the business venture must acquire, to upgrade its operational performance. The accuracy of the sizing summary generated by the tool depends on the input provided by the customers, i.e. more the inputs provided, better is the footprint created and lesser is the offset present in the output values. The following paragraphs give a short description of topology diagrams, performance graphs and infrastructure diagram respectively.

The topology diagram portrays the intended number of machines along with their configuration, needed to deploy the services and also indicates how many services should be deployed on each machine. If the computed number of machines comes out to be very large, the customer is recommended to switch to the next larger configuration. Following figure represents a topology diagram for a large system implementation. Assuming 10 services are to be deployed on a platform like AMX, the diagram would look like this:

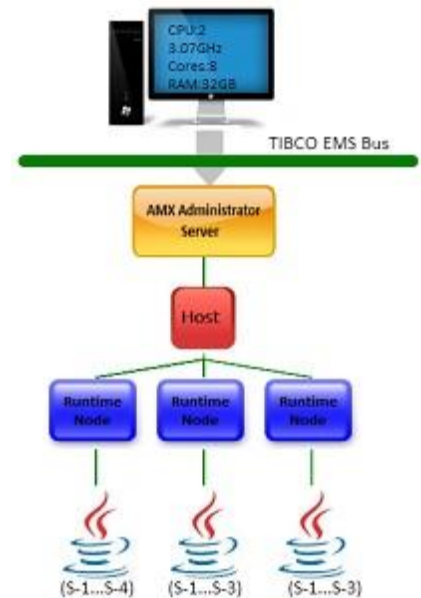

Fig 6. Topology diagram for AMX platform

Performance graphs play a very important role in depicting how the system is currently operating and to what extent it can continue to perform well. They also predict the point upto which the performance of the system could be scaled. Following graph displays how CPU utilization of a machine increases with increase in the number of services. The red region indicates that performance of the machine would degrade if the number of services deployed on that machine goes above a certain value, 12 in this case.

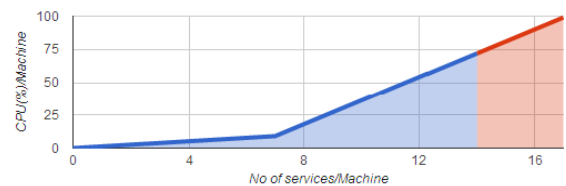

Fig 7. Performance graph

The infrastructure diagram gives a blueprint of how the various components required to establish the setup are connected to each other and how they communicate with one another. Following diagram shows an AMX specific infrastructure diagram.

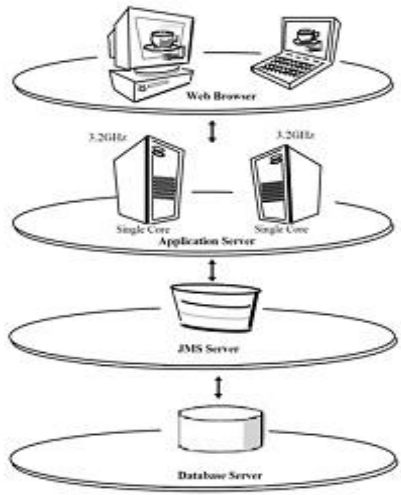

Fig 8. Infrastructure Diagram

## V. PRACTICAL APPLICATION

1. In an environment, where organizations are relying on SOA based solutions to solve their complex IT (Information Technology) issues, a Sizing Tool will act as a catalyst by determining the amount of hardware that will be needed to establish a stable and fault tolerant system. Having an automated and accurate tool to carry out the sizing process mitigates the risk of system breakdown and reduces the overall cost as well.
2. The sizing tool acts in accordance with the equations and algorithms to come up with real-time system configurations that can be put to effect.
3. This is a research oriented sizing solution being developed wherein the mathematical equations formulated to arrive at sizing requirements can be converted into a generalized mathematical model which then can be used to size other similar SOA platforms.
4. This web based tool is based on a generalized sizing API which resembles SaaS(Software as a Service), i.e. the same API can be extended to carry out sizing operations on other products and applications eliminating the overhead of implementing the algorithm from scratch.
5. In comparison to the currently available sizing tools in the market, this tool is build using a lightweight UI(User Interface) framework which makes it more reactive and user-interactive. It also provides end-to-end sizing solutions for deployment as well as runtime requirements.

## VI. CONCLUSIONS

In an era of fast changing business requirements, more and more SOA based platforms are emerging day-by-day. It is one of the architectural styles being adopted by the upcoming enterprises due to its key features such as service orientation, inter service communication via governance, underlying technology independence, etc. To make the most of this type of architecture, it is essential to carry out sizing of applications on such platforms to determine hardware requirements and performance details before actual implementation of the system. This not only makes the system resilient but also diminishes the failure costs that may arise due to dynamically changing business processes. The TIBCO sizing tool aims at giving an automated solution to the sizing problem with the help of extensible UI and core API.